\documentclass{PoS}

\title{Lattice ${\cal N}=4$ three-dimensional super-Yang-Mills}

\ShortTitle{3d super-Yang-Mills}

\author{\speaker{Joel Giedt}\thanks{JG and PM were supported in part by the Department of Energy, Office of Science, Office of High Energy Physics,
Grant No. DE-SC0013496. AL is supported by the Royal Society as a Royal Society University Research Fellowship holder.}\\
        Department of Physics, Applied Physics and Astronomy, Rensselaer Polytechnic Institute, Troy, New York, 12180 USA\\
        E-mail: \email{giedtj@rpi.edu}}

\author{Arthur Lipstein\\
        Department of Mathematical Sciences, Durham University, Durham, DH1 3LE, United Kingdom\\
        E-mail: \email{arthur.lipstein@durham.ac.uk}}

\author{Paul Martin\\
Department of Physics, Applied Physics and Astronomy, Rensselaer Polytechnic Institute, Troy, New York, 12180 USA\\
        E-mail: \email{martip8@rpi.edu}}

\abstract{We describe our recent work on the lattice formulation
of ${\cal N}=4$ three-dimensional super-Yang-Mills.  Our formulation
was based on the Donaldson-Witten twist, but we have also been
studying the formulation based on the Blau-Thompson twist by Joseph.
We find in the latter case there is a single counterterm necessary to
restore supersymmetry in the continuum limit, and that this counterterm
can be computed with a two-loop calculation in lattice perturbation
theory.  It is crucial that this three-dimensional model is
super-renormalizable.  We also describe some of the motivations for
studying three-dimensional theories, including mirror symmetry
and holographic cosmology.}

\FullConference{The 36th Annual International Symposium on Lattice Field Theory - LATTICE2018\\
		22-28 July, 2018\\
		Michigan State University, East Lansing, Michigan, USA.}

\newcommand{\mynote}[1]{}

\newlength{\dummysp}
\newcommand{\tr}{\mathop{{\hbox{Tr} \, }}\nolimits}

\newcommand{\beq}{\begin{eqnarray}}
\newcommand{\eeq}{\end{eqnarray}}
\newcommand{\nnn}{ \nonumber \\ }

\newcommand{\gappeq}{\mathrel{\rlap {\raise.5ex\hbox{$>$}}
{\lower.5ex\hbox{$\sim$}}}}
\newcommand{\lappeq}{\mathrel{\rlap{\raise.5ex\hbox{$<$}}
{\lower.5ex\hbox{$\sim$}}}}

\newcommand{\ben}{\begin{enumerate}}
\newcommand{\een}{\end{enumerate}}
\newcommand{\bit}{\begin{itemize}}
\newcommand{\eit}{\end{itemize}}

\newcommand{\cQ}{{\cal Q}}

\def\[{\left [}
\def\]{\right ]}
\def\({\left (}
\def\){\right )}

\begin{document}

\section{Motivations}
One of the motivations to study three-dimensional (3d) super-Yang-Mills (SYM) is mirror symmetry \cite{Intriligator:1996ex}.
This is a duality that relates physics on the Coulomb branch of a 3d gauge theory to the
physics on the Higgs branch of another 3d gauge theory, and {\it vice versa,} with Coulomb and Higgs
interchanged.  These branches have to do with
different vacua that are allowed within the theories, described by moduli space with a metric
and vacuum coordinates.  A nontrivial moduli space is indeed a common feature of supersymmetric
theories with extended supersymmetry, in this case ${\cal N}=4$ supercharges.
To be fair, the examples of mirror symmetry all involve SYM with additional matter fields,
whereas our lattice studies have so far focused on pure ``super-glue.''  However, it is
a first prerequisite to be able to study the gauge theory without matter, and then
once confidence has been built, to add the matter to further studies.

Another interesting feature of the 3d SYM theories is that the gauge coupling is dimensionful,
$[g^2] = 1$.  Thus the theory contains an intrinsic scale in the ultraviolet (UV) description.
Nevertheless, it is believed that in the infrared (IR), these supersymmetric gauge
theories flow to a nontrivial conformal field theory (CFT), without any fine-tuning of couplings.
This is interesting because the CFT must be free of scales.

Three-dimensional gauge theories with adjoint scalars, such as ${\cal N}=4$ SYM, are
of use in the scenario of holographic cosmology \cite{Afshordi:2016dvb}.
Indeed the SYM theory has exactly the type of quartic scalar interaction
that is needed.  It is claimed that holographic cosmology can do
a better job of modeling small angle ($\ell < 30$) cosmic microwave
background (CMB) than the conventional $\Lambda$CDM model.
Here, lattice simulations would predict large angle statistics
and anomalies---it is hoped.

Of course, 3d SYM would also be useful for studying the holographic
correspondence on $AdS \times X_6$, where $X_6$ is some compact manifold
that encodes the internal $R$ symmetry group of the gauge theory.

\section{Continuum theory}
The continuum theory may be obtained as a dimensional reduction of the 6d ${\cal N}=1$ SYM
to 3d.  Thus we begin with
\beq
{\cal L} = \frac{1}{2g^2} {\rm Tr} F_{\mu\nu} F_{\mu\nu}
+ \frac{i}{g} {\rm Tr} \Psi^T C \Gamma_\mu D_\mu \Psi
\eeq
The dimensional reduction proceeds rather naively:
\beq
A_\mu \to A_i, \quad \phi_\alpha, \quad
i = 0, 1, 2; \quad \alpha = 1, 2, 3
\eeq
for the gauge fields, and for the fermions
\beq
\Psi_p, \quad p = 1, \ldots , 8
\to
\psi_a^I, \quad a=1,2; \quad I = 1, 2, 3, 4
\eeq
Note that the index $I$ corresponds to the ${\cal N}=4$ of this SYM.

The topological twists of this theory, upon which the lattice formulations
are based, are most easily seen from the 6d $\to$ 4d dimensional reductions.
In that case we have the spacetime group
\beq
SO(4) \simeq SU(2)_l \times SU(2)_r
\eeq
We also have the internal R symmetry group
\beq
SU(2)_R \times U(1)_R
\eeq
From these, we derive the twisted rotation group
\beq
SU(2)' = {\rm diag} ( SU(2)_r \times SU(2)_R )
\eeq
Writing all of the fields in terms of representations under this group, the Lagrangian
takes the Donaldson-Witten form \cite{Witten:1988ze}
\beq
g^{2}\mathcal{L}_{4d}^{\mathcal{N}=2} &=& \tr \bigg(\frac{1}{4}\mathcal{F}_{\mu\nu}\mathcal{F}^{\mu\nu}+\frac{1}{2}\mathcal{D}_{\mu}\bar{\phi}\mathcal{D}^{\mu}\phi-\alpha\left[\phi,\bar{\phi}\right]^{2}
\\ && \qquad-\frac{i}{2}\eta\mathcal{D}_{\mu}\psi^{\mu}+i\alpha\phi\left\{ \eta,\eta\right\} -\frac{i}{2}\bar{\phi}\left\{ \psi_{\mu},\psi^{\mu}\right\} +\mathcal{L}_{\chi} \bigg),
\\
\mathcal{L}_{\chi} &=& {\rm Tr} \left(\frac{i}{8}\phi\left\{ \chi_{\mu\nu},\chi^{\mu\nu}\right\} -i\chi^{\mu\nu}\mathcal{D}_{\mu}\psi_{\nu}\right)
\eeq
The 3d theory is then obtained simply by the replacement
\beq
{\cal D}_2 \to [\phi_3, \cdot]
\eeq
This was the basis of our lattice formulation in \cite{Giedt:2017fck}.

An alternative, Blau-Thompson, twist \cite{Blau:1996bx} is used in \cite{Joseph:2013jya}.  
The dimensional reduction of the 6d rotation group to 3d is denoted by:
\beq
SO(6) \to SO(3) \times SO(3) \simeq
SU(2)_E \times SU(2)_N
\eeq
Then one takes the twisted rotation group to be
\beq
SU(2)' = {\rm diag} ( SU(2)_E \times SU(2)_N )
\eeq
This corresponds to
the earlier $\cQ=8$ formulation by orbifold method \cite{Cohen:2003qw}.

\section{Lattice formulation based on Donaldson-Witten twist}
What we found is that in order to preserve the exact nipotent $\cQ$, we must complexify everything.
We have a dynamical lattice spacing, as usual in these twisted/orbifold lattices.
We then lift the additional fields with generic mass terms.  The lattice Lagrangian is then:
\beq
&&
\mathcal{L} = {\rm Tr} \left(\frac{1}{4}\bar{\mathcal{F}}_{\mu\nu}(n)\mathcal{F}_{\mu\nu}(n)+\frac{1}{2}\bar{\mathcal{D}}_{\mu}^{+}\bar{\phi}(n)\mathcal{D}_{\mu}^{+}\phi(n)-\alpha\left[\phi(n),\bar{\phi}(n)\right]^{2}\right.
 \nnn &&
\left.+\frac{i}{2}\bar{\mathcal{D}}_{\mu}^{+}\eta(n)\psi_{\mu}(n)+i\alpha\phi(n)\left\{ \eta(n),\eta(n)\right\} -\frac{i}{2}\bar{\phi}(n)\left(\psi_{\mu}(n)\bar{\psi}_{\mu}(n)+\bar{\psi}_{\mu}\left(n-e_{\mu}\right)\psi_\mu\left(n-e_{\mu}\right)\right)\right)
 \nnn && +\mathcal{L}_{\chi},
\\
&&
\mathcal{L}_{\chi}={\rm tr}\left[\frac{i}{8}\left(\phi(n)\bar{\chi}_{\mu\nu}(n)\chi_{\mu\nu}(n)+\phi\left(n+e_{\mu}+e_{\nu}\right)\chi_{\mu\nu}(n)\bar{\chi}_{\mu\nu}(n)\right)\right.
\nnn
&&
\left.-\frac{i}{2}\left(\bar{\chi}_{\mu\nu}(n)\bar{\mathcal{D}}_{\mu}^{+}\bar{\psi}_{\nu}(n)+\chi_{\mu\nu}(n)\mathcal{D}_{\mu}^{+}\psi_{\nu}(n)\right)\right].
\eeq
Lattice gauge invariance and $\cQ$ invariance of
\beq
\chi_{\mu\nu}(n)=\frac{1}{2}\epsilon_{\mu\nu\rho\lambda}\bar{\chi}_{\rho\lambda}\left(n+e_{\mu}+e_{\nu}\right).
\eeq
implies
\beq
\sum_{\mu=1}^{4}e_{\mu}=0.
\eeq
So, the theory must be 3d.
After using the equations of motion,
\beq
&& \mathcal{L} = \cQ \,{\rm Tr} \left(\frac{1}{4}\chi_{\mu\nu}(n)\mathcal{F}_{\mu\nu}(n)+\frac{1}{2}\bar{\mathcal{D}}_{\mu}^{+}\bar{\phi}(n)\psi_{\mu}(n)+\alpha\eta(n)\left[\phi(n),\bar{\phi}(n)\right]\right)
\\ && -\frac{1}{8}\epsilon_{\mu\nu\rho\lambda}{\rm tr}\left(\mathcal{F}_{\mu\nu}(n)\mathcal{F}_{\rho\lambda}\left(n+e_{\mu}+e_{\nu}\right)\right),
\eeq
The last term is $\cQ$ invariant using the lattice Bianchi identity.
Note that
$\cQ$, which is nilpotent, acts as:
\beq
&& \cQ\,\phi(n)=0,\,\,\, \cQ\,\bar{\phi}(n)=i\eta(n),
\\
&&
\cQ\, \eta(n)=\left[\bar{\phi}(n),\phi(n)\right],\,\,\,
\\
&&
\cQ\,\mathcal{U}_{\mu}(n)=i\psi_{\mu}(n),\,\,\, \cQ\,\bar{\mathcal{U}}_{\mu}(n)=-i\bar{\psi}_{\mu}(n)
\\
&&
\cQ\,\psi_{\mu}(n)=\mathcal{D}_{\mu}^{+}\phi(n),\,\,\, \cQ\,\bar{\psi}_{\mu}(n)=\bar{\mathcal{D}}_{\mu}^{+}\phi(n)
\\
&&
\cQ\,\chi_{\mu\nu}(n)=\bar{\mathcal{F}}_{\mu\nu}(n)+\frac{1}{2}\epsilon_{\mu\nu\rho\lambda}\mathcal{F}_{\rho\lambda}\left(n+e_{\mu}+e_{\nu}\right).
\eeq

\section{Counterterms}
The fine-tuning to recover supersymmetry in this 3d theory is calculable because the theory is super-renormalizable.
We have found in our detailed investigations that all counterterms are one-loop or two-loop, and not higher.
So, there is a finite number of counterterms that can all be calculated in lattice perturbation theory.
Calculating the loop integrals must be done numerically, but it is doable since we are only in 3d.
Perturbative corrections include the effects of the mass terms\footnote{These
mass terms are absent in the Blau-Thompson twist advocated by Joseph.} and scalar $\cQ$ breaking.

As a reference point for restoring supersymmetry in a lattice theory,
there is a finite counterterm in supersymmetric quantum mechanics with na\"ive discretization \cite{Giedt:2004vb}.
Doublers appearing in the one-loop correction to the scalar propagator (cancelled by scalar loop in the $\cQ$-exact case)
give rise to this counterterm.  That earlier study also shows the power of Symanzik improvement
in terms of obtaining ``good supersymmetry.''

We are currently coding up the Joseph discretization because it is much cleaner.
It has only one counterterm due to the exact $\cQ$, point group and lattice gauge invariance.
The lattice Lagrangian is especially simple
\beq
\mathcal{L}(n)=\frac{1}{g^{2}} \cQ{\rm Tr}\left(\chi_{ab}(n)\mathcal{D}_{a}^{+}\mathcal{U}_{b}(n)+\eta(n)\bar{\mathcal{D}}_{a}^{-}\mathcal{U}_{a}(n)+\frac{1}{2}\eta(n)d(n)+B_{abc}(n)\bar{\mathcal{D}}_{a}^{+}\chi_{bc}(n)\right)
\eeq

For dimension counting it is best to have canonical normalization (otherwise everything is very confusing)
\beq
\Phi \to g \Phi, \quad \Psi \to g \Psi, \quad {\cal U}_m = \frac{1}{ag} e^{ag {\cal A}_m}
\eeq
After this is done, the dimensions of the fields are given by:
[boson] = 1/2, [fermion] = 1, [d]=3/2, [F]=3/2, [$\cQ$] = 1/2.
So since each loop is multiplied by factors of
$[g] = 1/2$
we must have no more than dim=2 to be unsuppressed by lattice spacing.
But the operators must also be fermionic if they are under $\cQ$, i.e., are $\cQ$-exact.
So restricting to $\cQ$-exact operators, they must be of form
\beq
\cQ {\rm Tr} \Psi, \quad \cQ {\rm Tr} \Psi \Phi, \quad
\cQ ( {\rm Tr} \Psi {\rm Tr} \Phi)
\eeq
where $\Psi$ and $\Phi$ generically represent fermions and bosons respectively.
Indeed we have also found that there are no closed operators that are relevant or marginal in this theory.\footnote{Details
will be provided in a forthcoming publication \cite{forthcoming}.}
Point group, lattice gauge invariance, limit to
\beq
\cQ {\rm Tr} \eta = {\rm Tr} d
\eeq
This has the effect of shifting the $d$ equation of motion by a constant.
Perturbative calculations are underway, including Symanzik improvement, which obviously involves
many higher dimensional operators.
E.g., at leading order in this improvement, we must write down all dimension three $\cQ$ invariant fermionic operators.

\section{Conclusions}
The holographic cosmology does not really require supersymmetry, 
but it will be interesting to see how supersymmetry impacts large angle predictions.
Concrete realization of Symanzik improvement versus supersymmetry in 3d should be very enlightening.
In the Blau-Thompson twist we only have one counterterm to determine in order to get full ${\cal N}=4$ 
supersymmetry.  We have found that the one-loop contribution vanishes
identically, but the two-loop contribution needs to be
performed numerically.  Symanzik improvement will require more counterterms
and diagrams, but will significantly improve the supersymmetry of the
lattice simulations.  It is worth mentioning that there is also a
suggestion of how to preserve all of the supersymmetry in this
3d SYM theory using a modified Leibnitz rule \cite{DAdda:2007hnx},
though we do not know how to realize such a proposal on a computer
because of the complexities of the braided fields.

\bibliographystyle{JHEP}
\bibliography{3dQ8_giedt_lipstein}

\end{document}